\documentclass[fleqn,twoside]{article}
\usepackage{espcrc2}

\usepackage{graphicx}

\newcommand{\AmS}{{\protect\the\textfont2
  A\kern-.1667em\lower.5ex\hbox{M}\kern-.125emS}}

\input{aliases}

\title{\Bc and Excited \B States --- A Tevatron Review}

\author{R.~K.~Mommsen\address{School of Physics and Astronomy at the University of Manchester\\Manchester, M13 9PL, UK}\address{Fermi National Accelerator Laboratory\\Batavia, IL 60510, USA}
on behalf of the CDF and \Dzero collaborations}

\begin{document}

\begin{abstract}

\vspace{1pc}
\end{abstract}

\maketitle

\section{INTRODUCTION}
In this paper recent results from the CDF and \Dzero experiments on heavy flavor spectroscopy are reported. Both experiments are using up to 1.1\invfb of data delivered by the Tevatron proton-antiproton collider at the Fermi National Accelerator Laboratory, Batavia, IL, USA.
The CDF and \Dzero detectors are described in references~\cite{Abe:1988me,Abazov:2005pn}.

\section{\boldmath PROPERTIES OF THE \Bc}
Although discovered in 1998 by CDF~\cite{Abe:1998wi}, the properties of the \Bc remain poorly measured due to small samples of candidates available until recently. In Run II of the Tevatron, CDF and \Dzero experiments have accumulated enough data to study the \Bc in greater detail. Being the last discovered ground state of the \B meson and the only meson with two heavy quarks of different flavor, the \Bc is a great laboratory for potential models, HQET, and lattice QCD. Its mass, lifetime, decay properties, and production are all of interest as many precise predictions have been made by theorists.

At the Tevatron, the \Bc is reconstructed in several decay channels containing a \jpsi meson. It is seen in the semileptonic modes $\Bc\to\jpsi\electron\nu X$ by CDF and in $\Bc\to\jpsi\mmu\nu X$ by \Dzero, as well as in the hadronic mode $\Bc\to\jpsi\pi$ by CDF.
The signal significance in all cases is over $5\sigma$. The semileptonic decays are used to determine the proper decay time of the \Bc~\cite{Abulencia:2006zu,Bc_lifetime:2004}:
\begin{tabbing}
CDF:    \=$\tau_{\Bc} = 0.474^{+0.073}_{-0.066}\pm0.033\ps$    \\[2mm]
\Dzero: \>$\tau_{\Bc} = 0.448^{+0.123}_{-0.096}\pm0.121\ps$
\end{tabbing}

Note, that only a fraction of available data is used by both experiments (CDF analyzed 360\invpb and \Dzero 210\invpb), so significant improvements of the measurements are expected in near future.
The measured values agree well with the theoretical prediction of $0.55\pm0.15\ps$ found in~\cite{Kiselev:2003mp}.

The hadronic mode $\Bc\to\jpsi\pi$ using the full data sample of 1.1\invfb yields the best mass measurement so far~\cite{Bc_mass:2006}. 
The selection criteria were tuned on the control sample $\B\to\jpsi\kaon$ to give a large signal while keeping the background low in order to improve the possibility of a significant \Bc observation. 
The $\jpsi\to\mumu$ candidates were formed using muon quality criteria and requiring the dimuon mass to be within 70\mevcc of the world average \jpsi mass.
The transverse momentum of the third track and of the \B candidate is used to distinguish between signal and background. The lifetime ($c\tau$) of the \B candidate needs to be positively displaced and have its uncertainty determined with good precision. A good fit of the combined vertex and of the \jpsi mass constrained fit is required. The \B candidate has to point to the primary vertex both in terms of a pointing angle and in terms of having a small impact parameter significance. For the third track, we require the impact parameter with respect to a secondary vertex determined by the $\jpsi\to\mumu$ candidate to be small and the impact parameter significance with respect to the primary vertex to be large.

\begin{figure}[t]
\begin{center}
\includegraphics[width=0.95\columnwidth]{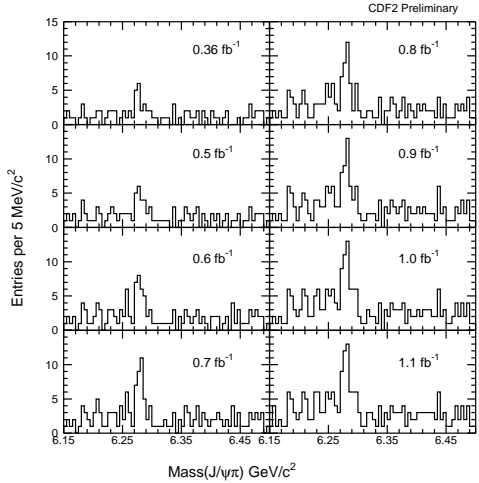}
\caption{$\jpsi\pi$ mass distribution for different integrated luminosities as data accumulated.}
\label{Bcmass_grow}
\end{center}
\end{figure}

\begin{figure}[t]
\begin{center}
\includegraphics[width=0.95\columnwidth]{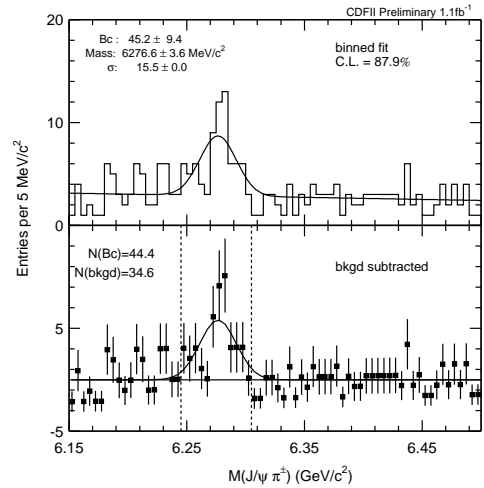}
\caption{$\jpsi\pi$ mass distribution in the 6.15 to 6.5\gevcc range with a superimposed Gaussian plus linear background binned fit (top). The $\jpsi\pi$ mass distribution with the linear background subtracted is shown (bottom) along with the number of events above background, $\mathrm{N}(\Bc)$ and the background in the 60\mevcc region between 6.245 and 6.305\gevcc.}
\label{Bcmass_fit}
\end{center}
\end{figure}

The selection criteria are tuned for a standard selection and a high-\pt selection. Both samples combined yield 11300 $\B\to\jpsi\kaon$ candidates with a small background of 250 events in the region between 5.4 and 5.5\gevcc. 
After fixing the selection criteria, the only change is the assignment of a $\pi$ versus \kaon mass hypothesis for the third track that is combined with the \jpsi.
Figure~\ref{Bcmass_grow} shows the number of \Bc candidates growing as a function of including additional integrated luminosity.
The top of Fig.~\ref{Bcmass_fit} depicts a binned fit using a linear background and a Gaussian signal shape for the \Bc data. The width of the Gaussian was fixed to the expected mass resolution $\sigma_R=15.5\mevcc$ observed in $\B\to\jpsi\kaon$ sample and scaled by a factor suggested by a Monte Carlo simulation. The number of \Bc candidates found  is $45.2\pm9.4$.
The bottom part of Fig.~\ref{Bcmass_fit} shows an unbinned fit to the data with the fitted background subtracted. 
In the region that is approximately $\pm 2\sigma_R$ wide between 6245\mevcc and 6305\mevcc, 44.4 \Bc candidates on a background of 34.6 are found.
In both cases the signal significance exceeds $6\sigma$, which has been confirmed by a toy Monte Carlo study.

The mass of the \Bc meson is determined by an unbinned log likelihood fit to a linear background and a Gaussian signal where the signal fraction, the background slope, and a scale factor for each event's mass resolution are fit parameters in addition to the mass. The scale factor for each event's mass resolution is fixed to 1.56 which is found from an unbinned log likelihood fit for the $\B\to\jpsi\kaon$ decay. The fit gives a mass of 
$m(\Bc) = 6276.5 \pm 4.0 \pm 2.7 \mevcc$,
which agrees well with the binned fit. The systematic error includes  uncertainties from the detector calibration, the tracking, and the small statistics fit procedure, which is the dominant contribution.
This result can be compared to the recent prediction using lattice QCD calculations: $m(\Bc) = 6304 \pm 12 ^{+18}_{-0}\mevcc$~\cite{Allison:2004be}.

\def\Bone      {\ensuremath{B_1}\xspace}
\def\Btwo      {\ensuremath{B_2^*}\xspace}
\def\Bss       {\ensuremath{B^{**}}\xspace}
\def\Bsone     {\ensuremath{B_{s1}}\xspace}
\def\Bstwo     {\ensuremath{B_{s2}^*}\xspace}
\def\Bsss      {\ensuremath{B_s^{**}}\xspace}

\section{\boldmath EXCITED \B STATES}
The spectroscopy of the $\overline{b}q$ system, where $q$ is either a $u$ or $d$ quark, is well understood theoretically.
The HQET describes a heavy-light state and predicts that there are four P-wave states, collectively called \Bss or $B_J$.
It is expected that two of them, $B_0^*$ and $B_1^*$, are wide states as they decay via S-wave. The other two states, \Bone and \Btwo, are narrow because they decay via D-wave. 
The quantitative understanding is not nearly as good. Few experimental
data are available on \Bss properties. 
However, since recently we are starting to see progress in this area.
Both CDF and \Dzero seek to observe and measure the two of the \Bss that have a narrow width, expected to be of the order of 10\mevcc. 
The other two P-wave states are ignored, as they are so wide that distinguishing them from combinatorial background is nearly impossible with the available data. \Bone decays only to $B^{*+}\pim$, while \Btwo can decay to either $B^{*+}\pim$ or the ground state $\Bp\pim$.

\begin{figure}
\begin{center}
\includegraphics[width=\columnwidth]{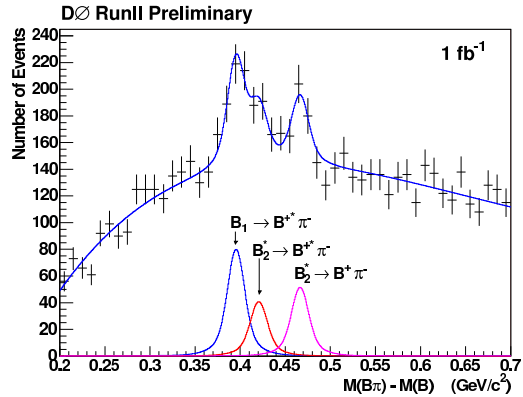}
\caption{The mass difference $m(\B\pi)-m(\B)$ for exclusive \Bss decays found by \Dzero. The line shows the fit using 3 Gaussians and a fourth-order polynomial as background function. The three signal peaks are shown separately.}
\label{Dzero:Bp_diff}
\end{center}
\end{figure}

The \Dzero experiment searches for all three decays of the narrow states mentioned above~\cite{excited_Dzero:2006}. 
The final state $\Bp\pim$ is reconstructed from $\Bp\to\jpsi\Kp$ where the \jpsi is found in the muon channel.
The photon coming from the decay of the excited state $B^{*+}\to\Bp\gamma$ is ignored. This leads to a shifted position of the mass peak for $\Bone\to B^{*+}\pim$ and $\Btwo\to B^{*+}\pim$.
The mass difference $m(\B\pi)-m(\B)$ for the $\Bp\pim$ candidates is shown in Fig.~\ref{Dzero:Bp_diff}. This is the first observation of separate peaks for the narrow \Bss states. \Dzero proceeds to fit this mass spectrum, assuming that
the widths of the two narrow resonances are the same and fixing the mass difference between the \Bstar and \Bp to 45.78\mevcc~\cite{PDG:2004}. The fit returns the masses and the width of these states:
\begin{eqnarray*}
m(\Bone) &= &5720.8 \pm 2.5 \pm 5.3\mevcc\\
m(\Btwo) - m(\Bone) &= &25.2 \pm 3.0 \pm 1.1\mevcc\\
\Gamma(\Bone) \doteq \Gamma(\Btwo) &= & 6.6 \pm 5.3 \pm 4.2\mevcc
\end{eqnarray*}
\Dzero also reports the production rates for these resonances:
\begin{eqnarray*}
\frac{\BR(\Btwo\to \Bstar\pi)}{\BR(\Btwo\to B^{(*)}\pi)} &= &0.513\pm0.092\pm0.115\\
\frac{\BR(\Bone\to B^{*+}\pi)}{\BR(\Bss\to B^{(*)}\pi)} &= &0.545\pm0.064\pm0.071\\
\frac{\BR(b\to\Bss\to\B\pi)}{\BR(b\to\Bp)} &= &0.165 \pm 0.024 \pm 0.028
\end{eqnarray*}

\begin{figure*}
\begin{center}
\includegraphics[width=\columnwidth]{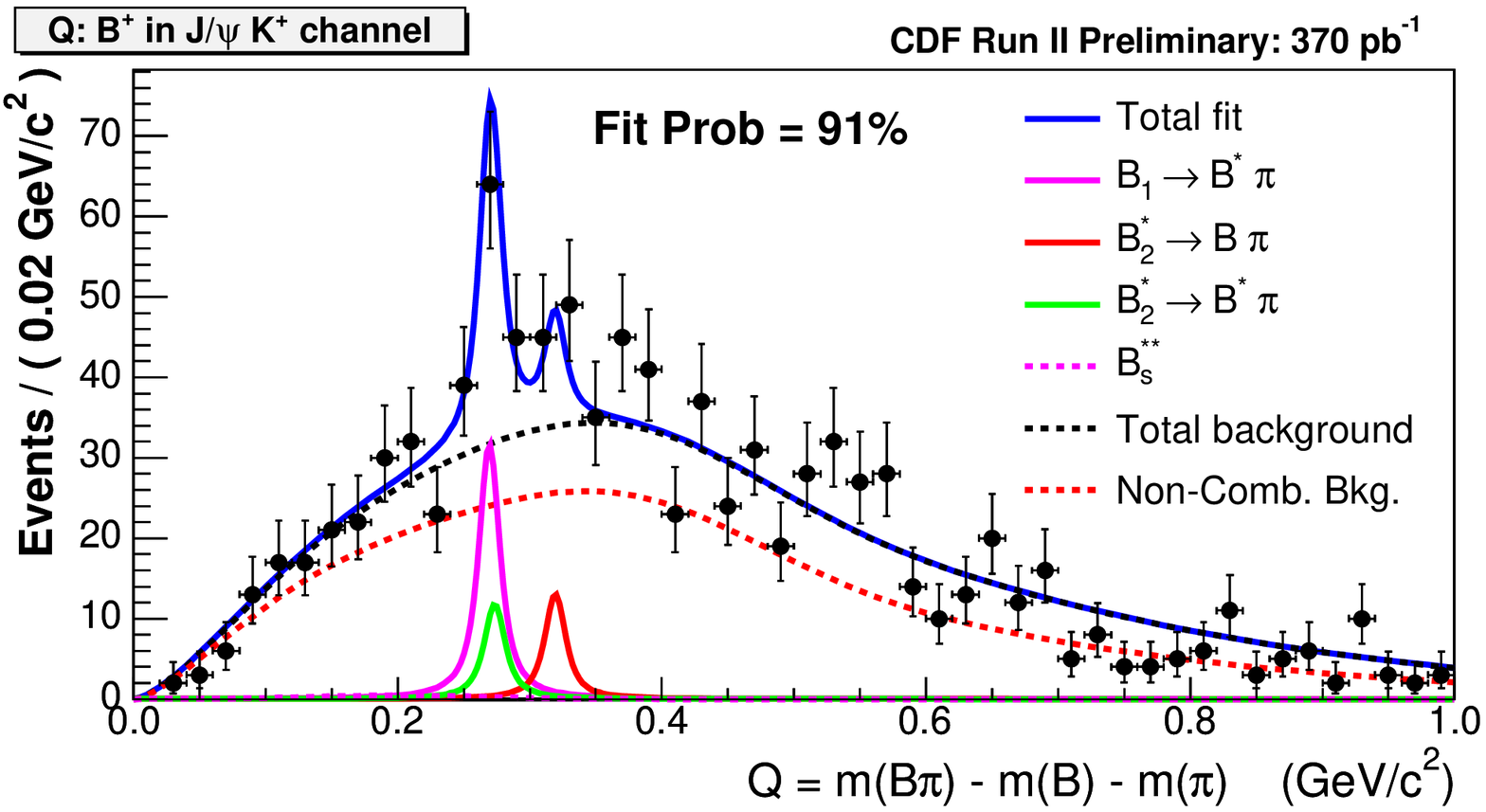}
\hfill
\includegraphics[width=\columnwidth]{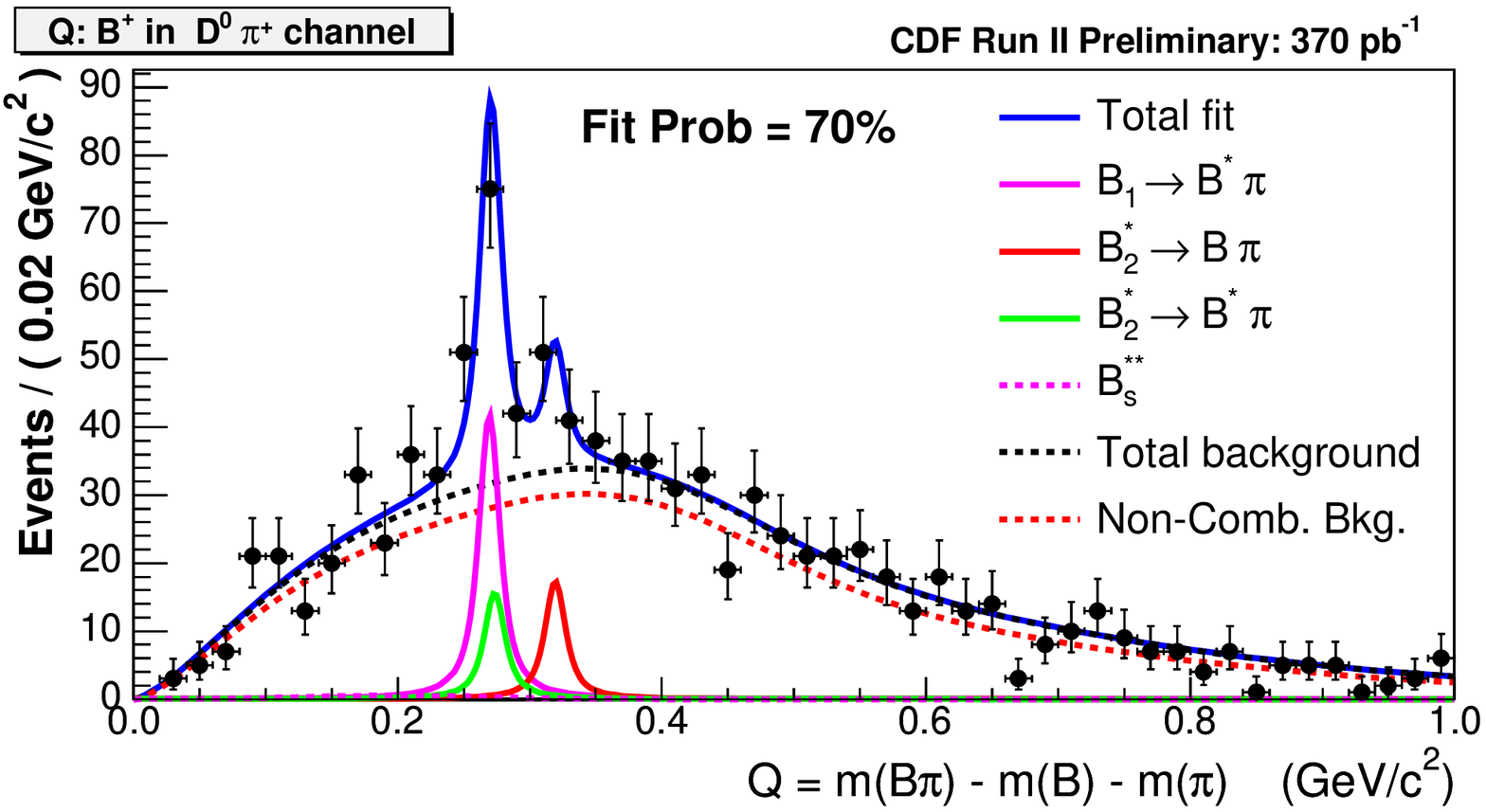}
\caption{Invariant mass difference for the \Bss candidates in the analysis from CDF. The fit shows the result of the simultaneous unbinned likelihood fit to the \Bss mass difference of the two samples $\Bp\to\jpsi\Kp$ and $\Bp\to\Dzb\pip$.}
\label{CDF:Bp_diff}
\end{center}
\end{figure*}

The CDF experiment performs a similar analysis~\cite{excited_CDF:2005}.
The same three decays of the \Bone and \Btwo are the subject of the measurement. 
The CDF sample of \Bp contains two signatures: $\Bp\to\jpsi\Kp$ and $\Bp\to\Dzb\pip$. The combined yield on 374\invpb of data is about 4000 signal candidates.
The mass difference $m(\B\pi)-m(\B)-m(\pi)$ for the reconstructed \Bss candidates is shown in Fig.~\ref{CDF:Bp_diff}. The fit of the mass spectrum is performed with the widths of both \Bone and \Btwo fixed to the theoretical expectation $\Gamma = 16 \pm 6\mevcc$~\cite{Falk:1995th}, and the ratio $\BR(\Btwo\to\B\pi) / \BR(\Btwo\to\Bstar\pi)$ is assumed to be $1.1 \pm 0.3$~\cite{excited_delphi:2004}. 
The result of the fit yields two mass measurements for \Bone and \Btwo which are not separated as in the case of \Dzero:
\begin{eqnarray*}
m(\Bone) &= &5734 \pm 3 \pm 2\mevcc\\
m(\Btwo) &= &5738 \pm 5 \pm 1\mevcc
\end{eqnarray*}

\section{\boldmath EXCITED \Bs STATES}

The heavy-light system $\overline{b}s$ is similar in its behavior to the $\overline{b}d$ systems. As well, the HQET predicts two narrow and two wide \Bsss states. These are even more difficult to study because of the lower production rates of \Bs mesons in comparison to more common \Bz and \Bp.
Due to the isospin conservation, the decay of $\Bsss\to\Bs\pi$ is highly suppressed. Thus the decay $\Bsss\to\Bp\Km$ is used. 

\begin{figure}
\begin{center}
\includegraphics[width=\columnwidth]{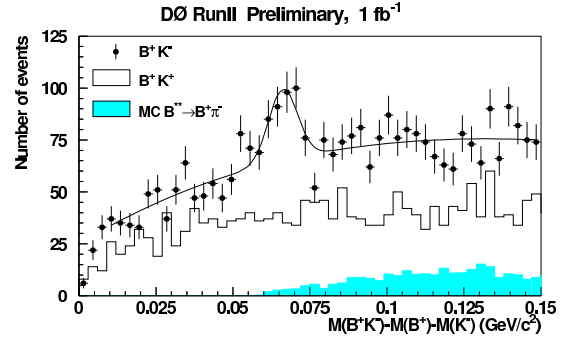}
\caption{The mass difference $m(\B\kaon)-m(\B)-m(\kaon)$ as measured by \Dzero. The smooth curve shows the fit of a third order polynomial representing the combinatoric background and a Gaussian representing the signal. The histogram shows the mass difference for $\Bp\Kp$ events. The solid histogram shows the MC distribution of the decay $\Bss\to B^{(*)}\pi$ where the $\pi$ is misidentified as a kaon.}
\label{Dzero:Bs_diff}
\end{center}
\end{figure}

\Dzero uses the same \Bp data sample as for the \Bss measurement. The invariant mass difference $m(\B\kaon)-m(\B)-m(\kaon)$ is shown in Fig.~\ref{Dzero:Bs_diff}. A clear peak is observed with a significance in excess of $5\sigma$. This peak is attributed to the process $\Bstwo\to\Bp\Km$. Thus, the mass of \Bstwo is $m(\Bstwo)=5839.1 \pm 1.4 \pm 1.5\mevcc$~\cite{Bs_excited_Dzero:2006}.

\begin{figure}
\begin{center}
\includegraphics[width=\columnwidth]{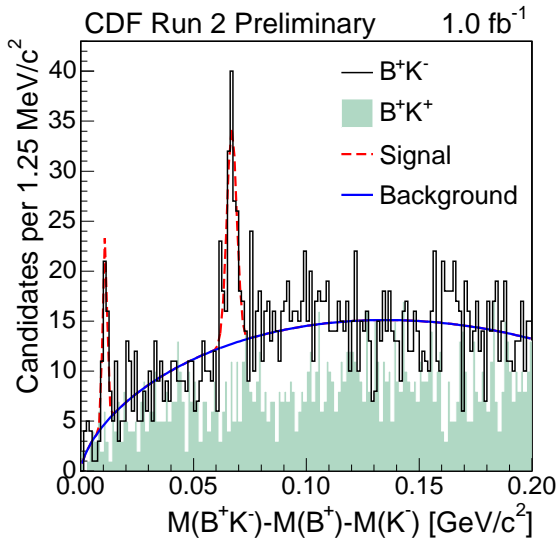}
\caption{The mass difference $m(\B\kaon)-m(\B)-m(\kaon)$ as measured by CDF. The line correspond to the projection of the unbinned maximum likelihood fit using both channels $\Bp\to\jpsi\Kp$ and $\Bp\Dzb\pip$ added together.}
\label{CDF:Bs_diff}
\end{center}
\end{figure}

CDF looks at the decays $\Bp\to\jpsi\Kp$ and $\Bp\Dzb\pip$ in 1.0\invfb of data. 
A total of 58k signal candidates are reconstructed using the decays $\jpsi\to\mumu$ and $\Dzb\to\Kp\pim$.
The invariant mass difference $m(\B\kaon)-m(\B)-m(\kaon)$ shown in Fig.~\ref{CDF:Bs_diff} has 2 distinct peaks. Both peaks have a significance in excess of $6\sigma$. Assigning the two peaks to the decays $\Bsone\to B^{*+}\Km$ and $\Bstwo\to\Bp\Km$, one finds:
\begin{eqnarray*}
m(\Bsone) &= &5829.4 \pm 0.2 \pm 0.6\mevcc\\
m(\Bstwo) &= &5839.6 \pm 0.4 \pm 0.5\mevcc
\end{eqnarray*}
This corresponds to a mass difference $m(\Bstwo)-m(\Bsone) = 10.20 \pm 0.44 \pm 0.35\mevcc$.

It is interesting to note that the masses for \Bone and \Bstwo found by CDF and \Dzero agree.
The mass difference in the $\overline{b}d$ and $\overline{b}s$ systems is expected to be very similar. This is indeed the case for the measurements done by CDF, albeit with large uncertainties. However, the mass difference $\Delta m \doteq m(\Btwo) - m(\Bone) = 25.2 \pm 3.0 \pm 1.1\mevcc$ as measured by \Dzero is significantly different from the ones found by CDF. 
Assuming the $\Delta m$ together with the \Bstwo mass measured by \Dzero, the \Bsone mass would be around 5814\mevcc. Thus, the mass would be too low for the decay $\Bsone\to B^{*+}\Km$, which would explain the absence of the second peak in the invariant mass difference $m(\B\kaon)-m(\B)-m(\kaon)$ (Fig.~\ref{Dzero:Bs_diff}).
This puzzle will hopefully be resolved in the future when analyses using higher statistics become available.

\section{CONCLUSIONS}
With over 1\invfb of data, many exciting results on heavy flavor physics are presently coming from the Tevatron experiments.
In this paper we have seen interesting results on heavy flavor spectroscopy.
The \Bc mass and lifetime is measured and agrees with the theoretical predictions.
The excited states \Bss and \Bsss offer an interesting laboratory to experimentally verify our understanding of quark interaction in bound states and to foster further development of non-perturbative QCD.
Overall, it is good time for flavor physics at the Tevatron as we are on the way to collecting multi-\invfb of data.


\end{document}